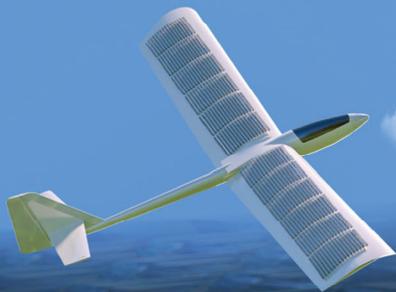
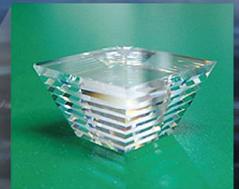
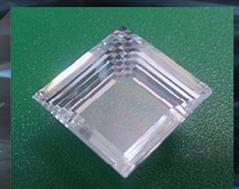
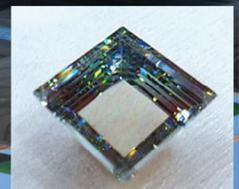
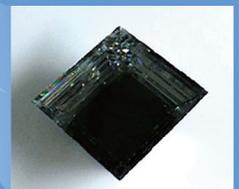
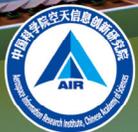

ARTICLE

Open Access

# Immersion graded index optics: theory, design, and prototypes

Nina Vaidya[1,2]✉ and Olav Solgaard[1]

**Abstract**
Immersion optics enable creation of systems with improved optical concentration and coupling by taking advantage of the fact that the luminance of light is proportional to the square of the refractive index in a lossless optical system. Immersion graded index optical concentrators, that do not need to track the source, are described in terms of theory, simulations, and experiments. We introduce a generalized design guide equation which follows the Pareto function and can be used to create various immersion graded index optics depending on the application requirements of concentration, refractive index, height, and efficiency. We present glass and polymer fabrication techniques for creating broadband transparent graded index materials with large refractive index ranges, (refractive index ratio)$^2$ of ~2, going many fold beyond what is seen in nature or the optics industry. The prototypes demonstrate 3x optical concentration with over 90% efficiency. We report via functional prototypes that graded-index-lens concentrators perform close to the theoretical maximum limit and we introduce simple, inexpensive, design-flexible, and scalable fabrication techniques for their implementation.

## Introduction

Harnessing the plentiful solar energy reaching the earth via photovoltaics will play a major role in satisfying our future energy needs in a sustainable way. One promising approach is concentrated photovoltaics[1], and several ways to achieve concentration are being used in the field[2,3]- Fresnel lenses[4,5], mirrors[6,7], parabolic concentrators[8,9], secondary high-index optics[10], waveguides[11–13], immersion lenses[14], surface nanotexturing[15]. The majority of these require active tracking of the Sun as they have to face the light source within a few degrees. Some of the above are passive concentrators, i.e., do not need to track the Sun, however they still offer modest acceptance angles that do not span the available 2π steradians. We present a passive concentrator device, AGILE (Axially Graded Index LEns)[16]. Henceforth, this immersion graded index optical concentrator is written as AGILE in the manuscript. AGILE does not need solar tracking and follows the cosine projection maximum limit, concentrating light incident on it from all angles.

The AGILE allows for non-pointing (i.e., removing the need to track the Sun) concentration systems which reduce the amount of photovoltaic (PV) material required but also efficiently absorb diffuse light. Light scattering is present due to cloud cover and atmosphere; and diffuse light can be as high as 20% even on a sunny day[17]. Figure 1a depicts how light is concentrated in the AGILE concentrator. Light rays incident from the entire 2π steradians enter the larger aperture with refractive index (RI) of 1, curve towards the normal via refraction along the height of the cone in the axial gradient RI, reflect from the sidewalls, and reach the smaller output aperture with high RI, e.g., silicon with RI ~ 3.5, without need to track the light source. Figure 1c portrays the AGILE concentrator array system, made up of the repeated unit shown in Fig. 1b, that absorbs all the incident light and hence appears dark. Video clip of the AGILE array system is attached in the supplementary material. In this video, the AGILE does not have metallic reflective sidewalls so that the graded index material can be visualised. AGILE allows near perfect antireflection and coupling, encapsulation, space for circuitry and cooling, and conformal design. These immersion graded index optics can also realize

Correspondence: Nina Vaidya (nina.vaidya@gmail.com)
[1]Electrical Engineering, Stanford University, Stanford, CA 94305, USA
[2] Faculty of Engineering and Physical Sciences, University of Southampton, Southampton SO16 7QF, UK

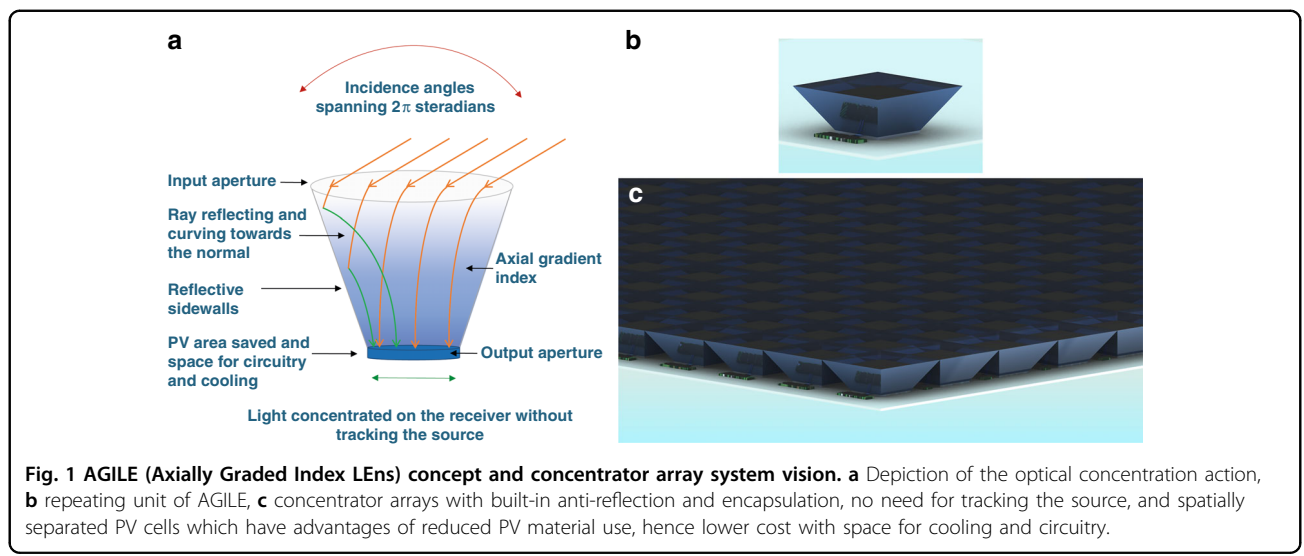

**Fig. 1 AGILE (Axially Graded Index LEns) concept and concentrator array system vision. a** Depiction of the optical concentration action, **b** repeating unit of AGILE, **c** concentrator arrays with built-in anti-reflection and encapsulation, no need for tracking the source, and spatially separated PV cells which have advantages of reduced PV material use, hence lower cost with space for cooling and circuitry.

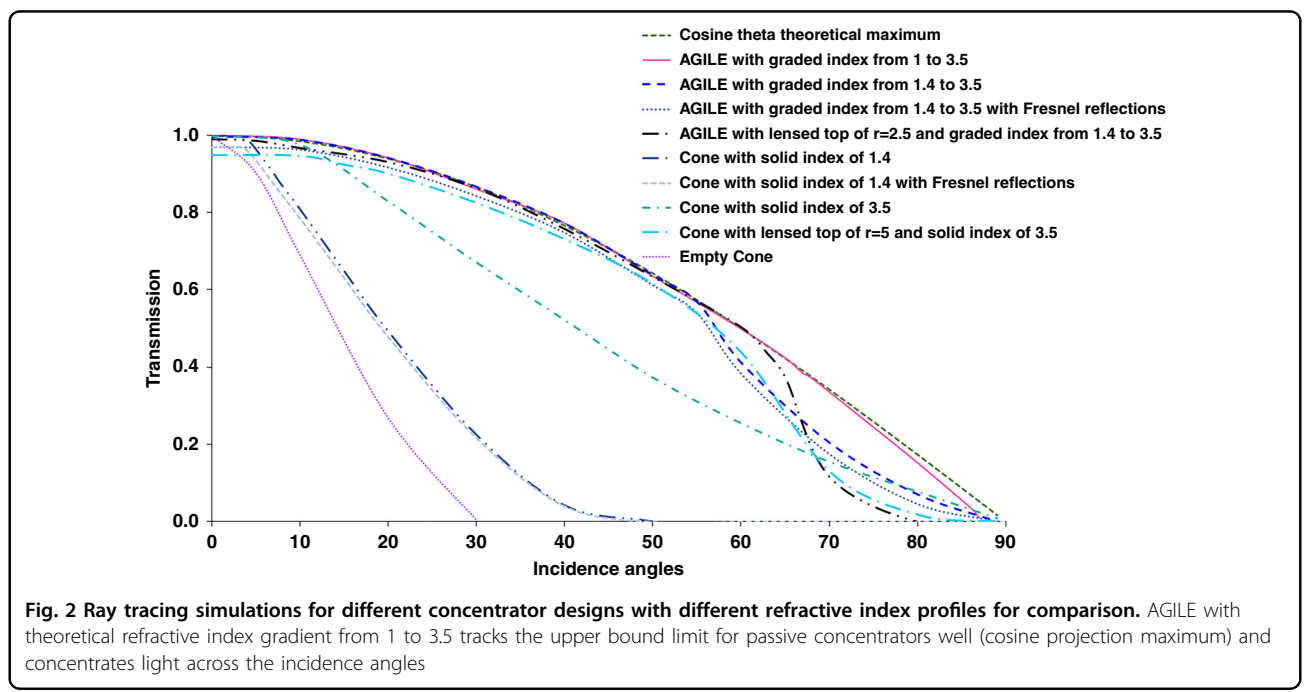

**Fig. 2 Ray tracing simulations for different concentrator designs with different refractive index profiles for comparison.** AGILE with theoretical refractive index gradient from 1 to 3.5 tracks the upper bound limit for passive concentrators well (cosine projection maximum) and concentrates light across the incidence angles

applications in areas like light management in solid state lighting, laser couplers, display technology etc.

## Comparison of different concentrator designs

Ray tracing simulations were performed with the software FRED. We performed simulations to compare optical concentration efficiencies of cone geometries with different RI profile fills. All the concentrators simulated in Fig. 2 have the same geometry- the input diameter is 3.5, the output diameter is 1, the height is 5, and the interior sidewalls of the cone are optically reflective. These are dimension-independent simulations in the ray domain, i.e., in the regime where the device dimensions are several times larger than the wavelength of the incident light. Details of scale invariance are presented in appendix A in the supplementary file. The angle of the incident ray array, theta, was swept from 0 to 90° in one plane as the structure has rotational symmetry. Figure 2 shows the cone geometries' optical concentration efficiency, i.e., light transmission to the smaller output aperture versus incidence angle compared to the cosine theta theoretical maximum, i.e., projection limit when not tracking the light source. The results show that the concentrator that is air filled, homogeneously filled with RI = 3.5, or homogeneous filled with RI = 3.5 along with a lens-top reject a substantial amount of the incident light. In contrast, an AGILE with a linear gradient index from ambient to the detector material (i.e., RI from 1 to 3.5) concentrates light close to the theoretical limit. In these simulations, Fresnel reflections at the top surface are not included in most curves unless denoted in the chart legend, as an antireflection thin film can be added at the top surface. However, as seen in the curves where Fresnel reflections are included, the AGILE is unchanged, while the lens-top and high-index filled cones have reduced transmissions.

### The AGILE theoretical concept

While designing optical concentrators, the constant brightness theorem[18–20] imposes strict limits. The theorem says that the optical power-flow per unit of area and solid angle cannot be increased through a passive optical system: luminance is invariant. Stated this way, the theorem is not complete and higher luminance (formerly described as brightness) can be achieved inside a high refractive-index (RI) material, as seen in optical immersion techniques in microscopy and lithography[21,22]. Mathematically, due to conservation of etendue in a perfect optical system, the concentration (C) in a 3-D optical concentrator can be expressed[23,24]:

$$C = \left\{\frac{a_{in}}{a_{out}}\right\}^2 = \{(n_{out} sin\emptyset_{out})/(n_{in} sin\emptyset_{in})\}^2 \qquad (1)$$

where $a_{in}$ and $a_{out}$ are the radii of the input and output apertures, $n_{in}$ and $n_{out}$ are the input and output refractive indices, and $\emptyset_{in}$ and $\emptyset_{out}$ are the input and output acceptance half angles. RI is denoted as n(z) in the equations. In solar concentrators, if we consider the input RI is unity and the output spread half angle as 90°, the equation reduces to $C = \{n_{out}/sin\emptyset_{in}\}^2$. Here, $\emptyset_{in}$ is close to 90°, so the concentration of the device:

$$C_{AGILE} = n_{out}^2 \qquad (2)$$

equation 1 states that optical concentrators accepting all incidence angles are not in violation of the constant brightness theorem, provided that the area reduction ratio is less than or equal to the square of the ratio of RI from input to output. In this paper, we show that concentration at this limit ($C = \{n_{out}/n_{in}\}^2$) can be achieved by gradually changing the RI from near unity in air to a high RI at the output. Extending Eq. 2, the RI continuously increases from input to output, while the area (A) of the concentrator decreases such that:

$$\int_{A(x,y,z_0)} n^2(x,y,z) dx dy = constant \qquad (3)$$

where x and y are the transversal coordinates, z is the axial coordinate, A is the cross sectional area, and n(x,y,z) is the RI. For a circularly symmetric structure of radius r(z) and a RI that is only a function of z (i.e., RI constant at each z plane), Eq. 3 can be written as

$$n(z) * r(z) = constant \qquad (4)$$

with Eq. 4 fulfilled, the number of electro-magnetic modes is constant along the height of the concentrator at each z plane; however, that condition in itself is not sufficient to ensure concentration. It is easy to find examples of structures that fulfill Eq. 4 but fail to concentrate light due to reflections from the sidewalls, e.g. jagged or discontinuous sidewalls or abrupt index variations. Therefore, the design challenge is to find concentrator shapes and corresponding index profiles that allows the electromagnetic modes of one layer to effectively couple to the next layer without unacceptable reflections.

Consider r(z) and n(z) as the sidewall and the index profiles respectively along z (height of AGILE). K is a non-zero constant and h is a dimensionless parameter that gives the height relative to the input radius, height = hK. $n_1$ and $n_2$ are the input and output indices, with boundary conditions: $r(0) = \frac{K}{n_1}, n(0) = n_1, r(hK) = \frac{K}{n_2}, \& n(hK) = n_2$. By using n(z)*r(z)=K (i.e., Eq. 4) and setting the curvature, i.e., second derivatives of r(z) equal to a constant ε, give the analytical solutions. By defining the second derivative as a constant, the solutions would have parametric continuity 'smoothness' of the order $ε^2$. Hence, this approach eliminates n(z) and r(z) that are unphysical and discontinuous. Different family of curves (i.e., complementary sidewall and index profiles) that create the AGILE design can be found. The analytical profiles for n(z) and r(z):

$$n(z) = \frac{2n_2 hK}{\left\{z^2 n_2 \epsilon h + z\left[2 - n_2 \epsilon K h^2 - 2\left(\frac{n_2}{n_1}\right)\right] + 2hK\left(\frac{n_2}{n_1}\right)\right\}}$$

$$r(z) = \frac{\left\{z^2 n_2 \epsilon h + z\left[2 - n_2 \epsilon K h^2 - 2\left(\frac{n_2}{n_1}\right)\right] + 2hK\left(\frac{n_2}{n_1}\right)\right\}}{2n_2 h}$$

$$(5)$$

We have mainly simulated and fabricated concentrators with linear sidewalls. For ε = 0, r(z) is linear and n(z) is hyperbolic:

$$n(z) = \frac{n_2 hK}{\left\{z\left[1 - \left(\frac{n_2}{n_1}\right)\right] + hK\left(\frac{n_2}{n_1}\right)\right\}}$$

$$(6)$$

$$r(z) = \frac{\left\{z\left[1 - \left(\frac{n_2}{n_1}\right)\right] + hK\left(\frac{n_2}{n_1}\right)\right\}}{n_2 h}$$

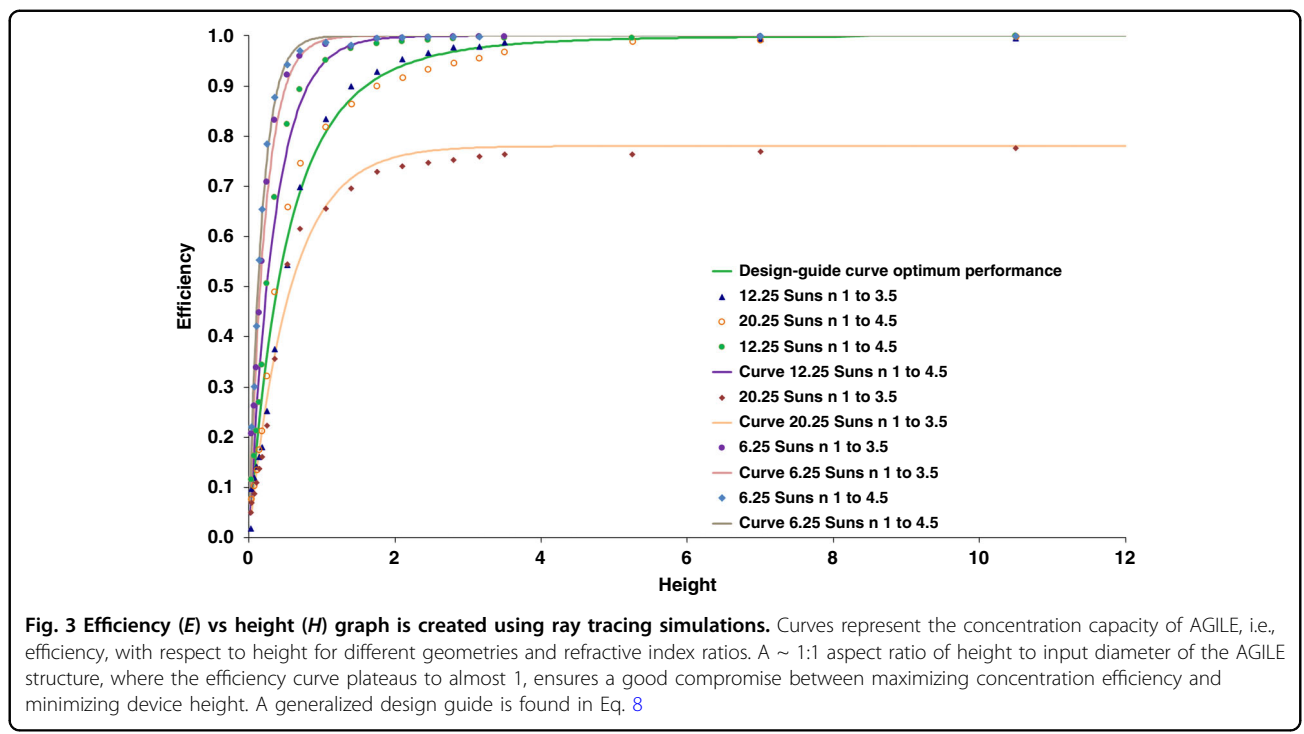

**Fig. 3 Efficiency (*E*) vs height (*H*) graph is created using ray tracing simulations.** Curves represent the concentration capacity of AGILE, i.e., efficiency, with respect to height for different geometries and refractive index ratios. A ~ 1:1 aspect ratio of height to input diameter of the AGILE structure, where the efficiency curve plateaus to almost 1, ensures a good compromise between maximizing concentration efficiency and minimizing device height. A generalized design guide is found in Eq. 8

**Height optimization**

The aim of the height optimization is to design a short device to save material and weight, while maintaining efficiency. AGILE device is scalable in the ray-domain (i.e., as long as the concentrator geometry is several times larger than the wavelength of light). Details in appendix A in the supplementary file. Scalable here means the concentrating phenomena of the AGILE works in the same way even if the geometry is increased or decreased by a scale factor. In other words, if we scale the Cartesian coordinates by a common factor, the ray paths are unaffected. The consequence of scale invariance is that the characteristics and therefore the performance of a particular AGILE is purely determined by the height to input diameter ratio. In appendix A, we show that: (1) reflections in the AGILE are scale invariant and (2) the ray tracing equation in the gradient index of the AGILE is scale invariant. Therefore, the height optimization study is also scalable. Shorter AGILEs are lossy, and efficiency increases as the AGILE gets longer. One of the main ray-rejection mechanisms is from the sidewalls of the top corner area of the AGILE. As the height of concentrator gets shorter, the corner angle gets smaller. The refractive index gradient inside the AGILE in the corner area is not large enough to bend the rays before they are reflected out by the sidewall. The combined effect of the small corner angle and not enough index variation in this space is that rays escape out of a short AGILE. A tall AGILE gives a large corner angle and hence near-perfect light capture. In reality the longer the AGILE gets, the larger the material transmission losses and greater the number of reflections/bounces; hence, there is a limit to how long the AGILE should be made. Another source of loss, apart from non-ideal height of the device, could be abrupt index variations. Reference[25] reports that in inhomogeneous media, even if the $\Delta RI$ (i.e., difference between largest and smallest RI) is large, as long as the RI slope is equal to $\Delta RI/(2\lambda)$ or smaller, the reflectivity is negligible. However, there is an abrupt limit around $\Delta RI/[(1/20)\lambda]$ where reflectivity increases sharply. In AGILE, the RI slope is 5 orders of magnitude lower than the abrupt limit. The index variation happens gradually over a large distance when compared to the wavelength in AGILE, so this abrupt limit, where reflections from index variations become significant, is not reached.

We use ray tracing simulations performed with the software FRED to find the optimum height as shown in Fig. 3. Both in simulations and in experiments, we focused on structures with linear sidewall geometry. Every curve in Fig. 3 is created from several simulations. Each point represents a particular AGILE device of a specific height, concentration, and index variation, and the point value gives the integration of transmission across 0° to 90° incidence angles compared to the input light, i.e., overall performance across all angles for a particular geometry. This process repeated for different heights made a curve for one concentration and index variation combination. Each curve is based on about 400 simulations. In Fig. 3, the upper bound is the maximum efficiency of 1 (all light incident on the input aperture reaching the output). The 'height to input diameter' ratio dictates the efficiency, as this ratio controls the sidewall angle and hence the retention of rays inside the structure. Accordingly to Eq. 3, the number of Suns concentration factor (i.e., the area reduction ratio) of 6.25, 12.25, and 20.25 are chosen to be less than, equal to, and greater than the square of index variation ($3.5^2$ and $4.5^2$) in the simulated AGILEs, to evaluate how these curves differ from each other. For all the simulations the input diameter is the square root of the concentration and output diameter is 1. For the 6.25 Suns AGILEs the input diameter is 2.5 and output diameter is 1, for the 12.25 Suns AGILEs the input diameter is 3.5 and output diameter is 1, and for the 20.25 Suns AGILEs the input diameter is 4.5 and output diameter is 1. Height is normalized with output diameter of 1. For the various curves in Fig. 3, the efficiency almost reaches the maximum of 1 when the height of the simulated device is approximately equal to the input diameter. For example, for the curve '12.25 Suns, n 1 to 3.5', a height of 3.15 with input diameter of 3.5, i.e., height : input diameter of 0.9, gives an efficiency of over 98%. A ~ 1:1 aspect ratio of height : input diameter of the AGILE structure, where the efficiency curve plateaus to almost 1, ensures a good compromise between maximizing efficiency and minimizing height.

Figure 3 curves represent the concentration capacity of AGILE with respect to height for different geometries and refractive index ratios. All the simulation data in Fig. 3 closely follow the Pareto function of the form:

$$E = E_{\max}\left(1 - \frac{1}{(1 + aH)^b}\right) \quad (7)$$

with a and b as constants, $H$ as height, and $E$ as efficiency that approaches its maximum value $E_{\max}$, which is the smaller of 1 and $M = \frac{n_{out}}{n_{in}} / \frac{r_{in}}{r_{out}}$, as height gets large, where $n_{in}$ and $n_{out}$ are the input and output indices, and $r_{in}$ and $r_{out}$ are the input and output radii. Here height is denoted as '$H$' and is different to the small 'h' parameter used before in Eq. 6. As described in Eq. 3, when the area reduction ratio, as in the concentration factor, is equal to the square of the index ratio, the AGILE is at its optimum performance. This optimum performance is shown as the 'design guide' fitted to the height optimization data for '12.25 Suns with n varying from 1 to 3.5' and '20.25 Suns with n varying from 1 to 4.5', where M = 1. The '12.25 Suns with n varying from 1 to 4.5', '6.25 Suns with n varying from 1 to 4.5', and '6.25 Suns with n varying from 1 to 3.5' curves lie above the design-guide curve and the '20.25 Suns with n varying from 1 to 3.5' curve falls below the design guide, as they respectively have more and less index gradation compared to the area reduction factor. The generalized design-guide curve is given as:

$$E = E_{\max}\left(1 - \frac{1}{\left(1 + 0.02\left(\frac{RM^2}{N}\right)H\right)^{40M}}\right) \quad (8)$$

where $M = \frac{n_{out}}{n_{in}} / \frac{r_{in}}{r_{out}}$, $N = \frac{1}{n_{in}} - \frac{1}{n_{out}}$, and $R = \frac{1}{r_{out}} - \frac{1}{r_{in}}$.

The above design guide characterizes how the efficiency, as in the concentrating action of the AGILE, changes with the height and can be used to create various immersion graded concentrators depending on the application requirements. The AGILE is scale invariant as described above. Therefore, the height optimization study is also scalable, and this generalized result in Eq. 8 can be applied to various geometrical concentrations and refractive index ratios, as these Pareto function curves fall above, below, or track the optimum performance.

### Practical AGILE design

For photovoltaic (PV) systems where the light enters the input aperture in air (RI ≈ 1) and is absorbed in a high-index PV material (e.g. silicon with a RI ≈ 3.5), theoretical passive concentration (i.e., capturing light without tracking the movement of the source) given by Eq. 1 is ($3.5^2/1^2$) =12.25. Achieving this level of passive concentration requires development of materials with broadband transparency and having low to high indices creating a large range of RI spread. In this section we discuss realistic AGILE designs as compared to the theoretical simulations presented so far, i.e., available broadband transparent optical materials with a wide range of indices, the trade off between concentration and the available graded index range in the AGILE design, other practical features include tile-able/tessellated input apertures to maximize space and light capture, and an optimum height design drawing on the results of the height optimization.

### Fabrication

Robust, broadband transparent, and inexpensive graded RI materials were critical to the success of the AGILE. We have designed and fabricated two variations of AGILE prototype: a pyramid of stacked glass flats, and a cluster of cones filled with polymer layers of different RI[26]. The fabricated pyramid with a square-cross section is a tile-able design and the rotational symmetric polymer array made by overlapping cone shapes has a tile-able/tessellated input aperture as well - going from hexagons at the input to circles at the output. Prototype designs are seen in Fig. 4a, b. After extensive material search, selection, and characterization, broadband transparent optical glasses were found to be available having index of 1.5 to 2, and broadband transparent UV curable polymers having an index of 1.46 to 1.625. Broadband transparent in this application means high optical transmission across the solar spectrum from about ~300 nm to beyond ~1200 nm. Graded index materials were fabricated as layers of different indices using the different glasses and the different polymers selected; and this was an approximation (verified via simulations) to the theoretical continuous gradient refractive index according to Eq. 3. For practicality of fabrication, we have used graded index layers. A hyperbolic refractive index profile matches the as-fabricated linear sidewalls of the devices as informed by Eq. 6. The refractive indices of different glasses and polymers in the various layers across the height of the devices fabricated are shown in Fig. 4c (glass pyramid) and Fig. 4d (polymer cluster) along with the matching theoretical continuous hyperbolic gradient index according to Eq. 6 for comparison.

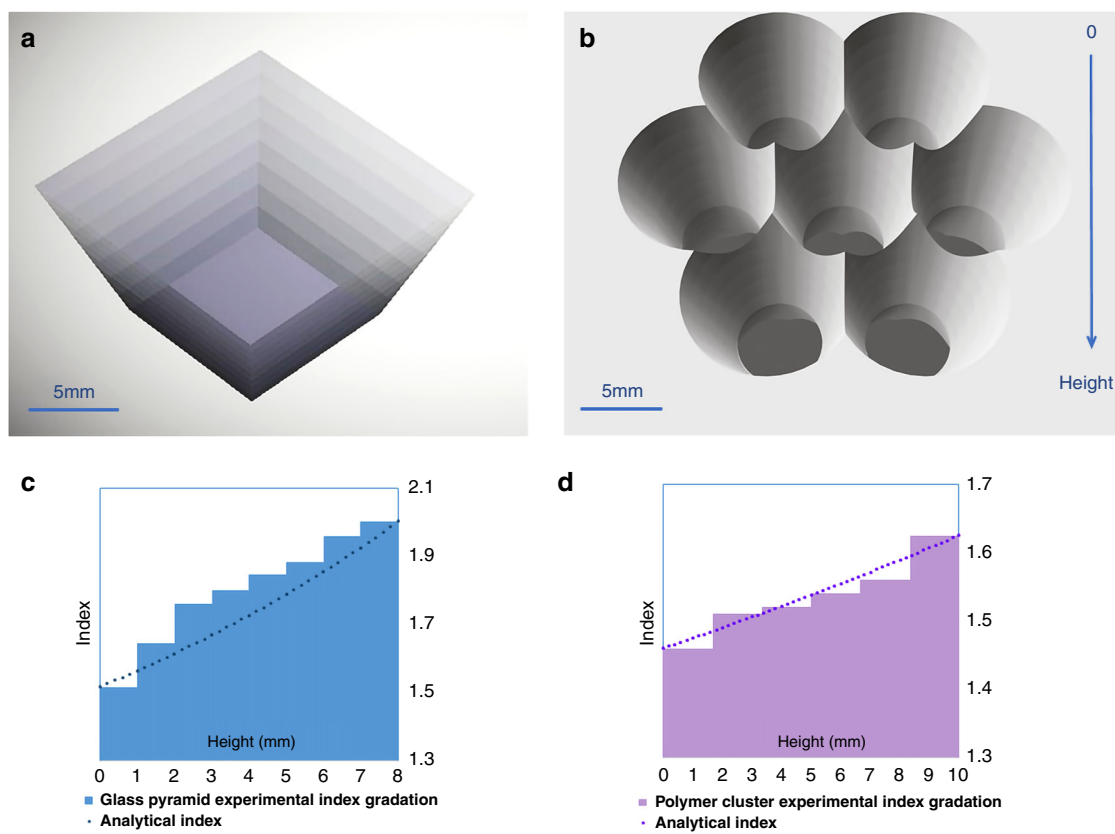

**Fig. 4 AGILE tile-able geometric designs and material selection of various different glasses and polymers having a large range of refractive indices along with high optical broadband transmission to create the graded index layers used in the prototypes.** AGILE structures with tile-able input surfaces: **a** pyramid with square cross section and **b** cluster of 7 overlapping cones. Height axis going from input aperture to output aperture is shown, which is true for both **a** and **b**. Refractive index gradation from low to high in the two structures along height of the device: **c** in the glass pyramid and **d** in the polymer cluster. The larger apertures at the top have a lower refractive index and the bulk material is graded to increase to a higher refractive index at the smaller apertures at the bottom. In graphs **c** and **d**, we compare the refractive indices of the different glasses and polymers used in the layers across the height of the fabricated AGILE devices (the steps) to the theoretical continuous hyperbolic index as per Eq. 6 (the dotted curves).

## Materials and methods
### Glass pyramid fabrication

The pyramid was made out of different optical glass flats, varying in index from 1.5 to 2 (list of glasses from Ohara Corp. in appendix B in the supplementary file). After an extensive search, these glasses were chosen from available optical-quality glasses, such that they have similar thermal expansion and glass transition temperatures, high broadband transmission in the solar spectrum, and RI evenly distributed in a wide range. To simplify fabrication, we create structures with linear sidewall geometry. The concentration efficiency difference between AGILE with the ideal hyperbolic index profile that matches a linear sidewall given in Eq. 6 and the layered index profile we fabricated is very small. In Fig. 4 c and d, we can compare how the selected material layers compare to the nominal hyperbolic index profile. The geometry of the pyramid was a square of side 14.5 mm down to a square of 8.5 mm giving a concentration of 3, along a total height of 8 mm with 8 glass layers, with each flat 1 mm thick. Glasses were cut into squares, polished on both sides, and rigorously cleaned. Bonding these different glass flats was a challenging task- melting glasses in a Kiln, silicate bonding[27], diffusion bonding in a hot press[28], anodic bonding[29], and bonding using SoG (Spin-on-Glass)[30] were attempted. We built a pressure vice for the glass flats using Teflon pins and a high-pressure mechanical vise. Direct bonding of the polished flats after plasma treatment and anodic bonding using a voltage (upto 30 kV) and heat were attempted. However, these attempts did not create the mechanically robust bonded glass stack that is necessary for machining. For the interfaces that were not anodically bonded, optical quality glue was used in between the glass flats to fill in any non-flat imperfections while being held in place under the pressure vise to remove excess glue, and this successfully created a robust graded index glass material layered from RI of 1.5 to 2. The fringe pattern before adding the glue was noted and was used to calculate the spacing between the non-ideal polished flats. There were wedges and curves seen in the fringe pattern which indicated non-uniform spacing and the maximum spacing was calculated to be 0.8 microns. The pyramid shape was micro-machined in this glass stack. To make the sidewalls reflective and not just dependent on total internal reflection, the pyramid's sidewalls were coated with aluminium. The pyramid's fabrication flow can be seen in Fig. 5.

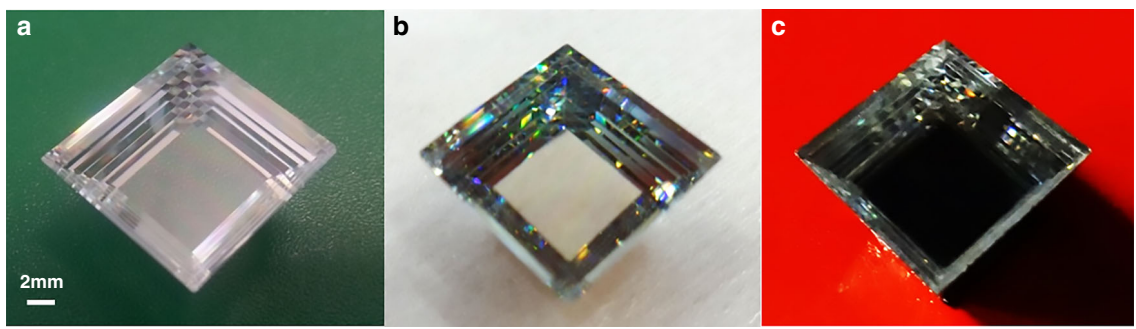

**Fig. 5 Fabrication steps of graded index glass pyramid concentrator.** Glass pyramid fabrication steps, **a** thin glass slabs of different refractive indices bonded together and the pyramid shape machined in the stack. Seen in the corners is the 'checkerboard' pattern, which is an optical illusion due to graded index layers, **b** aluminium deposited on the sidewalls, **c** pyramid in optical contact with solar cell absorbs and concentrates most of the incident light and appears dark.

### Polymer cluster fabrication

For the experimental demonstration of AGILE using polymers, broadband transparent UV curable optical polymers were chosen. The single AGILE structure had the issue that in order to collect all the light at the output, it had to be directly fabricated on or optically bonded to the photodetector with an appropriate anti-reflection coating. In a back-to-back AGILE, measurements of transmission were easier because the transmitted power was back in a low-index medium, i.e., air, so a photodetector with a standard AR (Anti-Reflection) coating was sufficient. In a back-to-back AGILE, light went from a larger aperture to smaller and then back to larger, thereby proving the concentration effect. The single AGILE was the basic structure to be demonstrated, the back-to-back AGILE was made for solely testing purposes and represents the ideal AGILE concentrator fabricated directly on a detector. An array structure was made – AGILE back-to-back array of 7 overlapping cones at the input that decrease from a diameter of 7 mm to 4 mm and increase back to 7 mm over a height of 10 mm having a concentration of 3 Suns (calculation of the over-lapping input aperture area in appendix C in supplementary file). Three structures are presented in Fig. 6c: 2 Suns AGILE cone (diameter of 7 mm to diameter of 5 mm, and index variation from 1.46 to 1.56), 3 Suns (7 overlapping cones of diameter of 7 mm to diameter of 4 mm) back-to-back AGILE with 10 layers (index variation from 1.46 to 1.56), and 3 Suns back-to-back AGILE with 12 layers variation (index variation from 1.46 to 1.625). The AGILE array is an extension of demonstration of the 2 Suns AGILE to 3 Suns (diameter of 7 mm to diameter of 4 mm). UV curable polymers having different indices were cured volumetrically layer by layer in polished aluminium cone shapes to create the AGILE prototypes[31]. Appendix D in supplementary file gives the fabrication steps used for creating these graded index polymer stacks.

## Discussion and results
### Characterization and results of the glass and polymer AGILEs

In characterizing the glass pyramid performance, care was taken to establish optical contact between the pyramid and the photodetector using an index matching liquid with RI of 1.7 (Cargille Labs). This RI value is not ideal but chosen because index matching liquids with higher values are corrosive to the high-index glass of the pyramid at the output and these liquids need to be handled as hazardous materials, or the liquids have colour, i.e., not broadband transparent. The fabricated AGILEs were tested by comparing the amount of laser light reaching the photodetector via a fixed aperture area (input aperture of AGILE) with and without the AGILE to evaluate the concentrator performance. The measurement set-up is depicted in Fig. 6a, which includes a red HeNe laser, beam expander, a rotational stage, a holder for AGILE, and a photodetector. Details of the measurement set-up and the test procedure are given in appendix E in the supplementary file.

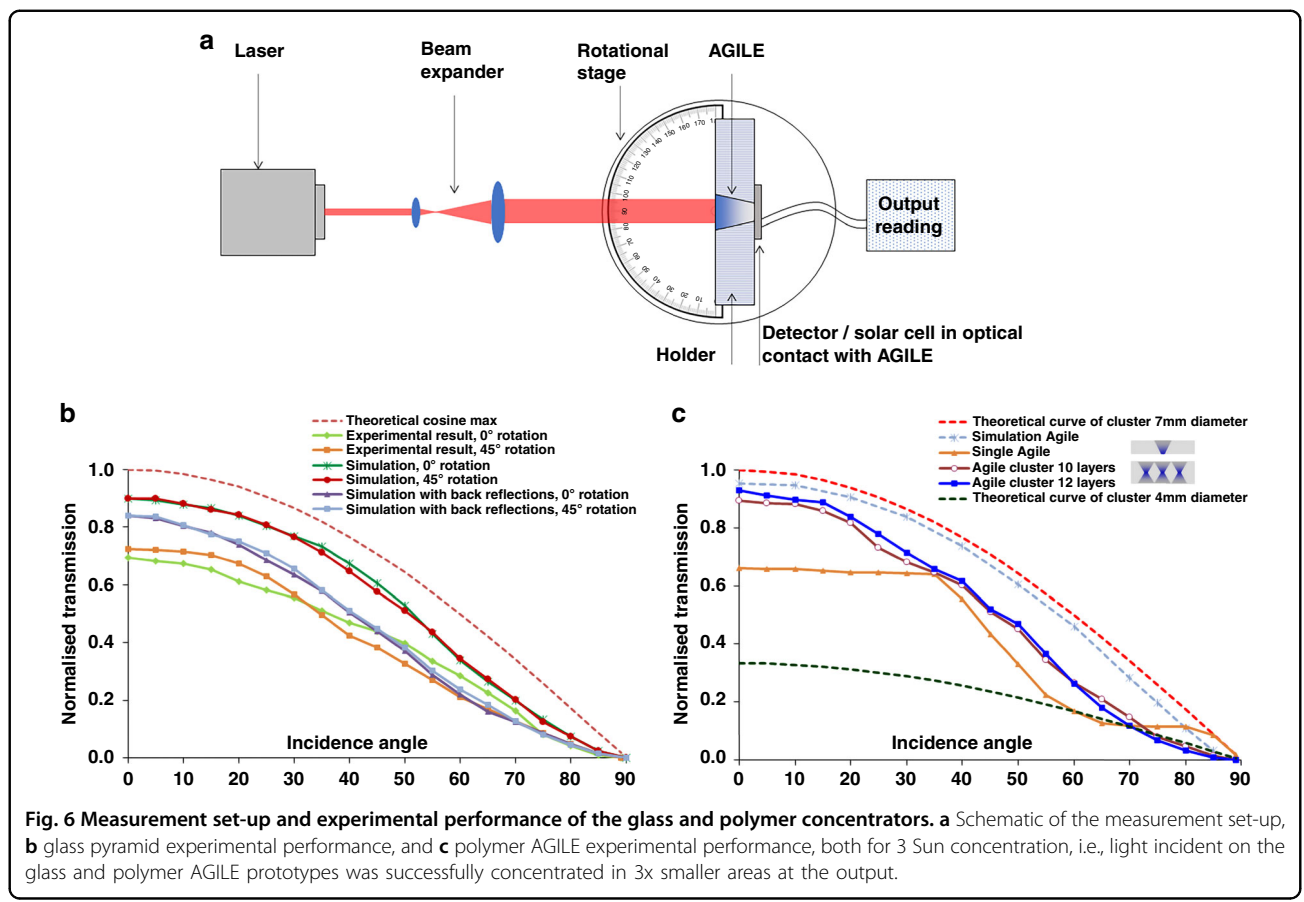

**Fig. 6 Measurement set-up and experimental performance of the glass and polymer concentrators. a** Schematic of the measurement set-up, **b** glass pyramid experimental performance, and **c** polymer AGILE experimental performance, both for 3 Sun concentration, i.e., light incident on the glass and polymer AGILE prototypes was successfully concentrated in 3x smaller areas at the output.

It can be seen from Fig. 6b that the two set of glass pyramid simulation results match in performance with the set of experimental results. There are, at least, two major distinct ways of measuring light concentration through a pyramid shape across different incidence angles. This is due to the fact that unlike the cone cluster the pyramid does not have rotational symmetry. The two set of results for the pyramid shape are (1) the angular measurements done with rotation along the side of square input aperture of the pyramid (annotated as '0° rotation' in Fig. 6b), and (2) the rotation along the diagonal of the square input aperture of the pyramid (annotated as '45° rotation' in Fig. 6b). The pyramid's 45° rotation optical transmission results are slightly lower than the 0° rotation results. This lower transmission can be accounted for by the corner reflection action of the pyramid which may increase the number of bounces of the rays, adding losses, and which is not present when only rotating/pivoting along the sides of the pyramid. Measurements were re-done and verified using green and blue LED sources and under a solar simulator and they track the results measured using the HeNe laser; proving broadband transmission and function as a solar concentrator. Ray tracing simulations of the pyramid concentrator performance are shown in Fig. 6b, together with experimental results. The losses at each intersection/interface in the layered structure were taken into account in the simulation by including the indices and thicknesses of the glass flats and the index matching layer. Light left the pyramid from the last glass layer of RI 2 then entered the 0.2 micron index matching liquid of RI 1.7, and then the silicon material of the detector with RI of about 3.5, in that sequence. The lower transmission performance of the pyramid as compared to the simulations was due to the fact that there are thin adhesive layers in between the glasses, the material loss of the structure, and the coatings on the solar cell detector, which were not accounted for in the simulations. At normal incidence, the simulation predicts a transmission of about 0.83. In comparison, the highest transmission experimentally measured at normal incidence through the pyramid is about 0.72.

As seen in Fig. 6c, the back-to-back polymer AGILE structures gave very good performance and were able to concentrate most of the light that was incident on the 7 mm diameter circular aperture through the smaller 4 mm diameter aperture half way along the axis in each cone of the cluster. The experimental concentration efficiencies should have been similar for the single AGILE and for the glass pyramid when compared to the back-to-back results, but reflections at the AGILE-photodetector interface led to reduced transmission, as seen in Fig. 6c. This lower transmission highlights the importance of optical and mechanical bonding and immersion between the AGILE and the detector. An ideal AGILE system includes fabrication of the AGILE directly on top of a solar cell detector, for perfect light capture, propagation, and conversion. The curves in Fig. 6c show a roughly sinusoidal modulation on the transmission through the AGILE. This harmonic modulation was more pronounced with a single wavelength than with a broadband illumination, typical of interference effects. To estimate the layer thickness d in this Fabry-Perot type resonance effect, $2\pi m = \frac{4\pi n_f d \cos\theta}{\lambda}$ with the RI of the layer, i.e., $n_f = 2$, m = integer = 1, $\lambda$ = 632.8 nm, and $\theta = 0°$; we find d = 157 nm, which is about the approximate antireflection coating/passivation layer thickness on a standard solar cell/detector. Constructive interference at the transmission side is due to the different resonant wave-fronts that arrive in phase, enhancing the signal at specific incidence angles and opposite for destructive interference dips. This also means that the harmonic variation is not caused by the graded index layers in the back-to-back AGILE, where layers of different types of polymers over a height of 20 mm make each layer 2 mm thick in the 10 layer cluster and 1.67 mm thick in the 12 layer one. As expected, the transmission curves through all AGILE devices fall between those of the conical clusters' 4 mm diameter (output aperture) and the 7 mm diameter (input aperture) theoretical maximum curves (red and green dotted lines). It was noteworthy that the back-to-back AGILE results (blue line with square symbols) follow the theoretical maximum cosine projection curve quite well over the full angular range (for example, at normal incidence, experimentally measured transmission through the polymer cluster is ~ 0.93, i.e., over 90% efficiency). The results demonstrate that light incident from all angles on the AGILE cluster was successfully concentrated in 3x smaller area.

Even though the inspiration for AGILE design did not come from nature, there are features of AGILE that can be found in the retina of fish (e.g. Gnathonemus) and compound eyes in insects (e.g. Lepidoptera), where a gradient index is present as anti-reflection to maximize transmission as well as to enable camouflage[32,33]. The human eye lens is also a layered structure of gradient RI ranging from about 1.406 to 1.386[34], i.e., has (RI ratio)$^2$ of 1.03. We have taken the gradient immersion index idea further and designed and fabricated devices with (RI ratio)$^2$ of up to 2, pushing the limit seen in nature, the fiber optics industry, or research[35–37].

## Conclusions

Immersion-graded-index optics as an effective non-tracking optical concentrator was conceptualized, simulated, and fabricated. The design choice of 1:1 aspect ratio of height to input diameter of the AGILE structure was found to ensure a good compromise between maximizing light-capture and minimizing the device height. We present a generalized design-guide equation relating the refractive indices and the geometry, which can be used to create various immersion-graded optical concentrators.

Search for appropriate broadband transparent materials and innovation of several fabrication techniques with multiple iterations were needed to create defect free materials with a large graded RI range. Approximating the ideal gradient index with a discrete stepped index yields results that are close to the theoretical maximum. The AGILE prototypes: the glass pyramid made by stacking of various different glass flats and the polymer array of overlapping cones made in an aluminium stencil, experimentally demonstrated a passive concentration of 3 Suns. The simple to test and verify, back-to-back AGILE array tracked the cosine theta theoretical maximum across all incidence angles. Difference between the results of single and back-to back devices brought home the importance of optical contact between the concentrator and the detector/solar cell, i.e., immersion. More sophisticated AGILE designs involve the incorporation of lens-top surface to increase light collection; sub-wavelength nano-structuring, porosity, and aerogels to create the low index side;[38–43] top surface strengthening for exposure to environment;[44] functionalized nano-particle filler layers[45] and nano-structuring[46,47] with passivation in the PV cell to create the high values of the RI, i.e., to increase the range of graded index; an optimized sidewall profile matching the index profile used according to Eq. 4; and a 3D gradient index in both axial and radial directions to decrease height of the concentrator and to eliminate the need of a reflective sidewall.

Successful optical quality fabrication and demonstration of concentrators using polymers: (1) enable lightweight-design-flexible structures and ability to fabricate directly on textured solar cells/detectors, (2) provide effective encapsulation and inexpensive panel packaging along with the PV costs offset by the optical concentration, and (3) allow possibility of large-scale manufacturing using spray coating, auto pipetting, multiscale imprint, casting, molding, and 3D printing[48]. Results of the functional prototypes demonstrate that immersion graded index technology can improve the way we concentrate and couple light many fold. The AGILE has the potential to greatly improve opto-electronic systems by reducing cost, increasing efficiency, providing a scalable concentration system with built-in anti-reflection and encapsulation without the need for tracking.

**Acknowledgements**
Fabrication work was done in the flexible clean room in Spilker, Stanford University, and we are thankful for valuable discussions and help with custom fabrication setups from Thomas E. Carver. We are grateful to Prof. Reinhold H. Dauskardt for his advice on material science. Thanks to Michael J. Mandella for help with FRED (optical simulation software used to design AGILE). We thank Timothy R. Brand for the fabrication help in the crystal shop. Thanks to Evan Scouros for work on ray trajectory theory. We are thankful for research discussions with J Provine on fabrication, Skip Huckaby on anodic bonding, Prof. Robert S. Feigelson for advice on hot pressing of glasses, and Kiarash Zamani Aghaie for help with optical mode theory. We thank GCEP (Global Climate and Energy Project) for funding and special acknowledgement to Stanford DARE (Diversifying Academia, Recruiting Excellence) fellowship. Part of the work was performed at the Stanford Nano Shared Facilities (SNSF), supported by the National Science Foundation award ECCS-2026822.

**Author contributions**
Original work of Nina Vaidya with guidance from PhD adviser Olav Solgaard.

**Conflict of interest**
The authors declare no competing interests.

**Supplementary information** The online version contains supplementary material available at https://doi.org/10.1038/s41378-022-00377-z.

# Supplementary Material for 'Immersion Graded Index Optics: Theory, Design, and Prototypes'

Nina Vaidya
Assistant Professor
Faculty of Engineering and Physical Sciences
University of Southampton
Southampton, SO16 7QF, UK
nina.vaidya@gmail.com
n.vaidya@soton.ac.uk

Olav Solgaard
Professor, Electrical Engineering
Stanford University
Stanford, CA 94305, USA
solgaard@stanford.edu

## Appendix A: Scale Invariance of AGILE

Acknowledgement: scale invariance draws on Evan Scouros's: Rays in linearly graded medium

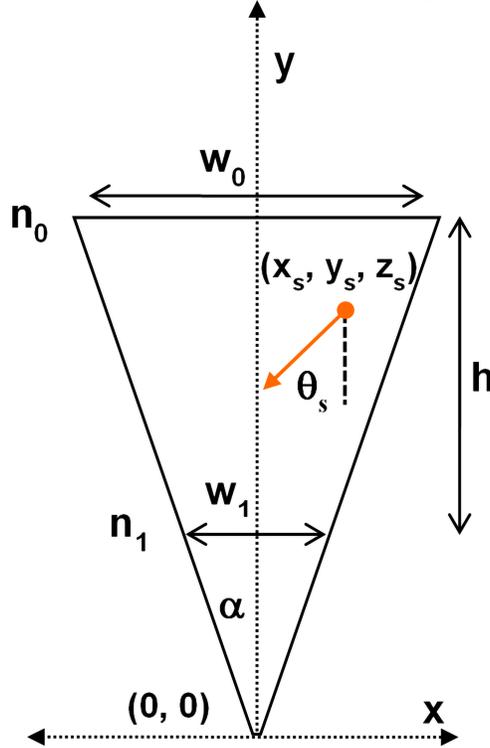

Figure A: AGILE geometry ($n_1 > n_0$)

We hypothesize that the AGILE is scale invariant in the sense that if we scale the coordinates by a common factor, the traces of all rays are unaffected. To prove this hypothesis, we must show that following requirements are fulfilled: (1) reflections in the AGILE are scale invariant and that (2) the ray tracing equation for the AGILE (i.e. in a gradient index) is scale invariant.

We will use the parameters defined in Fig. A that shows a cross section of the AGILE through its optical axis. The basic AGILE is rotationally symmetric around its optical axis, but AGILEs with more complex transversal shapes can also be used. Skewed rays will behave differently in AGILEs with different transversal shapes, but all shapes are scale invariant as will become clear.

**Basic Relationship:**
The reflection invariance follows directly from the shape of the AGILE. The indices and widths of the input and output apertures are related as follows

$$w_1 \cdot n_1 = w_0 \cdot n_0 \qquad 1$$

In other words, the gradient index is varied such that the index*width product is constant

$$n \cdot y \cdot 2 \tan \alpha = n_1 w_1 = n_0 w_0 \Rightarrow n = \frac{n_1 w_1}{y \cdot 2 \tan \alpha} = \frac{n_0 w_0}{y \cdot 2 \tan \alpha} \qquad 2$$

The slope of the sidewalls are given by the height and the difference in the widths

$$tan(\alpha) = \frac{\frac{w_0}{2} - \frac{w_1}{2}}{h} = \frac{w_0 - w_1}{2h} = \frac{\frac{n_1}{n_0}w_1 - w_1}{2h} \Rightarrow$$

$$\therefore tan(\alpha) = \frac{n_1 - n_0}{2h \cdot n_0} w_1 \qquad 3$$

Equation 3 proves the first of the requirements. The slope of the sidewalls, and therefore the reflection angles, are unchanged under a linear scaling in all three dimensions.

It is convenient to rewrite the index in the following form

$$n = \frac{n_1 w_1}{y \cdot 2 \tan \alpha} = \frac{n_1 w_1}{y \cdot 2 \left[\frac{\frac{n_1}{n_0}w_1 - w_1}{2h}\right]} = \frac{n_1 n_0 h}{y \cdot (n_1 - n_0)} \qquad 4$$

**Ray equation in gradient index:**

We require that Snell's law applies everywhere:

$$n \sin \theta = n_0 \sin \theta_0 = \text{Constant} = P \qquad 5$$

where $\theta$ is the angle that the ray makes with the optical axis. The starting point of the ray is at $x_s, y_s, z_s$, and the starting angle of the ray is designated $\theta_s$. Our treatment is valid for all points of origin of the ray.

Snell's law tells us that the path of the ray is given by

$$\tan \theta = \frac{d\sqrt{x^2 + z^2}}{dy} = \frac{1}{\sqrt{x^2 + z^2}} \frac{xdx + zdz}{dy} = \frac{1}{\sqrt{x^2 + z^2}} \frac{xdx}{dy} \frac{d\theta}{d\theta} \frac{dn}{dn} + \frac{1}{\sqrt{x^2 + z^2}} \frac{zdz}{dy} \frac{d\theta}{d\theta} \frac{dn}{dn}$$

$$\therefore \tan \theta = \frac{1}{\sqrt{x^2 + z^2}} \frac{xdx}{d\theta} \frac{d\theta}{dn} \frac{dn}{dy} + \frac{1}{\sqrt{x^2 + z^2}} \frac{zdz}{d\theta} \frac{d\theta}{dn} \frac{dn}{dy} \qquad 6$$

From Snell's law

$$\frac{dn}{d\theta} = \frac{d}{d\theta}\left[\frac{P}{\sin \theta}\right] = -P \frac{\cos \theta}{\sin^2 \theta} \qquad 7$$

From the equation for the gradient index (Eq. 4)

$$n = \frac{n_1 n_0 h}{y \cdot (n_1 - n_0)} \Rightarrow \frac{dn}{dy} = -\frac{n_1 n_0 h}{y^2 \cdot (n_1 - n_0)} \qquad 8$$

Now we arrive at the ray equation for the AGILE

$$\frac{\frac{xdx}{d\theta} + \frac{zdz}{d\theta}}{\sqrt{x^2 + z^2}} \left(-\frac{\sin^2 \theta}{P \cdot \cos \theta}\right)\left(-\frac{n_1 n_0 h}{y^2 \cdot (n_1 - n_0)}\right) = \tan \theta \Rightarrow \frac{\frac{xdx}{d\theta} + \frac{zdz}{d\theta}}{\sqrt{x^2 + z^2}} = \frac{Py^2 \cdot (n_1 - n_0)}{n_1 n_0 h \cdot \sin \theta} \Rightarrow$$

$$\int_{x_s, z_s}^{x, z} \frac{xdx + zdz}{\sqrt{x^2 + z^2}} = \frac{Py^2 \cdot (n_1 - n_0)}{n_1 n_0 h} \int_{\theta_s}^{\theta} \frac{d\theta}{\sin \theta} \qquad 9$$

$$\int_{x_s, z_s}^{x, z} \frac{xdx + zdz}{\sqrt{x^2 + z^2}} = \frac{Py^2 \cdot (n_1 - n_0)}{n_1 n_0 h} \left[\ln \frac{\sin \theta}{1 + \cos \theta} - \ln \frac{\sin \theta_s}{1 + \cos \theta_s}\right] \qquad 10$$

Using Snell's law to express the ray angle in terms of the y coordinate and the initial conditions, we find the desired ray equation:

$$\sin\theta = \frac{n_s \sin\theta_s}{n} = \frac{n_s \sin\theta_s}{\frac{n_1 n_0 h}{y \cdot (n_1 - n_0)}} = \frac{y \cdot (n_1 - n_0) \cdot n_s \sin\theta_s}{n_1 n_0 h} \qquad 11$$

$$\therefore \int_{x_s, z_s}^{x, z} \frac{xdx + zdz}{\sqrt{x^2 + z^2}} = \frac{Py^2 \cdot (n_1 - n_0)}{n_1 n_0 h} \left[ \ln \frac{\frac{y \cdot (n_1 - n_0) \cdot n_s \sin\theta_s}{n_1 n_0 h}}{1 + \sqrt{1 - \left(\frac{y \cdot (n_1 - n_0) \cdot n_s \sin\theta_s}{n_1 n_0 h}\right)^2}} - \ln \frac{\sin\theta_s}{1 + \cos\theta_s} \right]$$

$$\int_{\frac{x_s}{h}, \frac{z_s}{h}}^{\frac{x}{h}, \frac{z}{h}} \frac{\frac{x}{n_0 w_0} \cdot d\left(\frac{x}{h}\right) + \frac{z}{h} \cdot d\left(\frac{z}{h}\right)}{\sqrt{\left(\frac{x}{h}\right)^2 + \left(\frac{z}{h}\right)^2}} = P\left(\frac{y}{h}\right)^2 \frac{(n_1 - n_0)}{n_1 n_0} \left[ \ln \frac{P\left(\frac{y}{h}\right) \frac{(n_1 - n_0)}{n_1 n_0}}{1 + \sqrt{1 - \left(P\left(\frac{y}{h}\right) \frac{(n_1 - n_0)}{n_1 n_0}\right)^2}} - \ln \frac{\sin\theta_s}{1 + \cos\theta_s} \right] \qquad 12$$

Equation 12 shows that the second requirement for scale invariance is fulfilled. As the AGILE is linearly scaled in all dimensions, all rays with the same initial conditions ($x_s$, $y_s$, $z_s$, $P$, $sin\Theta_s$) follow the same path in the similarly scaled x,y,z coordinates. Furthermore, we see that the scaled ray path is only dependent on the scale factor given by the height ($h$), the initial position, and the angle of the ray.

**Appendix B**

| Ohara Glass | Refractive index at 589 nm | Linear expansion 10-7/K |
|---|---|---|
| S-BSL 7 | 1.517 | 75 |
| S-TIL 1 | 1.548 | 79 |
| S-TIM27 | 1.644 | 87 |
| S-LAM55 | 1.762 | 71 |
| S-LAM66 | 1.801 | 79 |
| S-NPH53 | 1.847 | 74 |
| S-LAH58 | 1.883 | 68 |
| S-LAH79 | 2.003 | 60 |

Table B: Glasses with different refractive indices selected such that they have broadband transparency in the solar spectrum i.e., high optical transmission across the whole solar spectrum say from ~300nm to beyond ~1200nm, and similar thermal expansion and glass transition temperature so that they are compatible once bonded with each other in a stack (data and glasses supplied by Ohara Inc.).

**Appendix C: Overlapping Conical Array Concentration Ratio**

We calculate the exact area ratio, i.e., geometric concentration of the cluster array fabricated (input area / output area). There is difference between the nominal design as seen in Fig. C (a) and what was fabricated as seen in Fig. C (b). This difference was the result of over-cutting by the reamer as it pulled extra metal with it when machining the conical shape from the fragile metal island in between the overlapping cones at the top input surface. This was anticipated in the design and fabrication trials; and hence what was fabricated was a tile-able/tessellated structure with almost no input aperture area wasted. The CAD design of seven cones going from a radius of 3.5mm to a radius of 2mm with a spacing of 6.35mm between the centers, became seven cones of 3.6mm input radius with the same output radii and spacing.

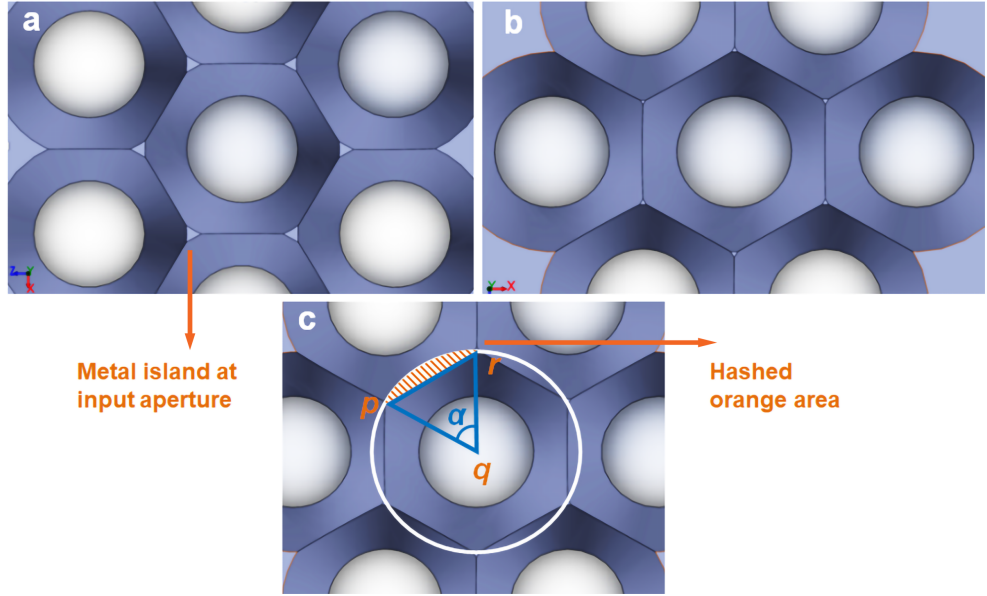

Figure C: Polymer cluster aerial view to evaluate the top surface area, (a) what was designed and (b) what was fabrication and (c) annotated figure for area calculations

We want to calculate the overlapping area, hashed orange area in Fig. C (c), in the top surface of the cones in order to calculate the total input aperture area. In this polymer cluster design, there are 7 overlapping cones which create 7 hexagon holes in the 2D plane of the top surface. To calculate the overlapping area, we will consider each arc sector and the enclosed triangle.

In the sector $pqr$ in Fig. C (c), angle subtended by arc pqr = $\alpha = 0.9816\ rad$

Area of sector $pqr = \frac{r^2\theta}{2} = 6.3611 mm^2$

Area of triangle $pqr = \frac{bh}{2} = 5.3875 mm^2$

Overlapping area between two neighbouring sectors = 2 x hashed orange region = 2 x (Area of sector $pqr$ - Area of triangle $pqr$)
$$= 2(6.3611 - 5.3875) mm^2$$
$$= 1.9471 mm^2 \approx 1.95 mm^2$$

12 such overlapping areas for a cluster of 7 circles $= 12 \times 1.95 mm^2 = 23.40 mm^2$

Total input aperture area = area of 7 circles – overlapping area calculated above
$= (7 \times \Pi \times 3.6^2) - (23.40) = 261.61 mm^2$

Concentration ratio = input area calculated above / area of 7 smaller circles at the output $= \frac{261.61 mm^2}{(7 \times \Pi \times 2^2) mm^2} = 2.97$

Therefore, concentration of the cluster after fabrication = 2.97

**Appendix D**
Creating a graded index stack using polymers:
1. Material search was done across various types of optically transparent polymers: silicones, acrylate polymers, polyimides, and polyurethanes (heat and/or UV curable resins). Ellipsometer and spectrophotometer measurements were done for various optical polymer film samples made of a fixed thickness to characterize the properties by measuring the film transmission and RIs across the solar spectrum. Polymers with broadband transparency (i.e., high optical transmission across the whole solar spectrum say from ~300nm to beyond ~ 1200nm) and having refractive indices evenly distributed in as large an index range as possible were chosen.
2. UV curable optical polymers from Norland Products with RIs 1.46, 1.51, 1.52, 1.54, 1.56, and 1.625 were chosen for fabricating the AGILEs. The single AGILE and 10 layer cluster were made using polymers with index 1.46, 1.51, 1.52, 1.54, and 1.56. The RI=1.625 layers, which require curing in an inert atmosphere (glovebox) was used in the 12-layer cluster fabricated.
3. Molds were made by reaming cone shaped cavities in aluminium metal plates. These cavities were polished using decreasing grit size sandpapers and polishing agents to make them optically reflective. After cleaning steps, a flat PDMS (Polydimethylsiloxane) film was attached using water soluble glue at the edges at the bottom of the AGILE reflective mold to seal the output in order to start filling in the graded index polymer layers (the UV curable optical polymers do not stick to PDMS and at the end of fabrication the device can be peeled off from the PDMS substrate). If curing is done first from the top there is the issue of uncured gel trapped below a cured top crust. To ensure a complete cure from the base, curing was first done with UV light incident through the PDMS layer from below the device, which was placed on a stand. . Later the cure was finished off from the top.

4. The UV cure parameters were tuned, such as, power and wavelength of the source, distance from the source, and the duration of cure. Some layers were fabricated in a glovebox due to need for inert atmosphere during the cure to avoid yellowing/clouding in air, e.g., oxygen inhibition in some of the polymers . Air gaps and bubbles were removed pre-cure by intermediate vacuum treatment steps. Some polymers shrunk during the cure and some needed age hardening after cure to achieve the required transparency and RI. These material differences were taken into account to make uniform and broadband transparent layers. Compatibility of the polymers with their neighbouring layers in the stack sequence was also tested before final fabrication.

5. Layers were filled volumetrically using pipettes to have a fixed layer height in order to complete the conical geometry. Each layer was filled and cured in several steps to ensure thin layers and hence a complete cure. UV curable polymer layers were formed one at a time starting from low-index to high-index polymers in order to fill half of the back-to-back shape; this process was followed by filling the upper part of the AGILE with high-index to low-index polymers. The single AGILE was filled from high-index to low-index. This completed the multi-stage polymer deposition and curing steps to create a graded index material with a large index variation.

**Appendix E**
Characterizing AGILE using HeNe laser

1. Test set up for measuring the light concentration ability included a red laser (HeNe Laser, 632.8 nm, 0.5 mW), beam expander, and a rotational stage as seen in Fig. 6a in the manuscript. Using a 3D printed holder the AGILE was fixed flat on the solar cell/photodetector which was mounted on a rotational stage. The AGILE was secured on the photodetector with both mechanical and optical bonding. 3D printed holders were painted black so that stray light is absorbed and only light going through the AGILE reaches the detector. The beam expander was used so we over fill/illuminate the input aperture uniformly. For accuracy of the measurements, the center of the input plane of the AGILE, i.e., input aperture was fixed at the center of rotation of the stage. The measurements have full 360º symmetry (important for solar applications) for axially symmetric shapes of the AGILEs tested like the over lapping polymer cluster. For the pyramid which is not axially symmetric two measurements were taken- rotation done along the side (0°) and rotation done along the diagonal (45°) of the square input aperture of the pyramid.

2. A visible wavelength laser is an ideal source to test the optical concentration efficiency of AGILE as it provides control over the incidence angle and also uniform area coverage at the input aperture. The current-voltage readings from the photodetector circuit were measured at different incidence angles (0 to 89°) of light through the AGILE to represent the light concentrated at the output. These values were compared to the readings without the AGILE but with the same input aperture on the photodetector to create normalized transmission curves. This allowed us to evaluate the effectiveness of AGILE by comparing (a) light collected by AGILE from a fixed input aperture area onto a 3x smaller photodetector area at the output; with (b) light collected on the same fixed input aperture area on the photodetector without AGILE.  The results graph is drawn from average values, each current-voltage reading was taken several times and the mean values were calculated. To make the results as directional as possible, measurements were taken with background lights off and the AGILE was placed at a fixed distance away from the source.